\documentclass[a4paper,11pt]{article}
\usepackage{pos}

\let\oldbibliography\thebibliography
\renewcommand{\thebibliography}[1]{%
  \oldbibliography{#1}%
  \setlength{\itemsep}{0pt}%
}

\title{Energy Loss in Small Collision Systems}

\author[a]{Coleridge Faraday}
\author*[a]{W. A. Horowitz}

\affiliation[a]{Department of Physics, University of Cape Town, Private Bag X3, Rondebosch 7701, South Africa}

\emailAdd{frdcol002@myuct.ac.za}
\emailAdd{wa.horowitz@uct.ac.za}

\abstract{We present leading hadron suppression predictions in $Pb+Pb$ and $p+Pb$ collisions from a convolved radiative and collisional energy loss model in which partons propagate through a realistic background, and in which the radiative energy loss receives a short pathlength correction. We find that the short pathlength correction is small for $D$ meson $R_{AA}(p_T)$ in both $Pb+Pb$ and $p+Pb$ collisions.  However the short pathlength correction leads to a surprisingly large reduction in suppression for $\pi$ mesons in $p+Pb$ and even $Pb+Pb$ collisions, providing a qualitative explanation for the rapid rise in $\pi$ meson $R_{AA}(p_T)$ at the LHC. We find that the size of the short pathlength correction to $R_{AA}(p_T)$ is acutely sensitive to the chosen distribution of scattering centers in the plasma. Furthermore we find that conventional elastic energy loss models, which apply the central limit theorem to the number of scatterings, dramatically overpredict suppression in $\mathrm{p}+\mathrm{A}$ collisions, calling for short pathlength corrections to elastic energy loss.}

\FullConference{HardProbes2023\\
 26-31 March 2023\\
 Aschaffenburg, Germany\\}

\begin{document}
\maketitle

\section{Introduction}

The modification of high-$p_T$ particle spectra provides crucial insights into the many-body dynamics of QCD in high-energy collisions. Significant suppression of high-$p_T$ particles in $A+A$ collisions has been observed at RHIC and LHC experiments \cite{PHENIX:2001hpc,STAR:2003pjh,PHENIX:2006ujp,PHENIX:2006mhb}, attributed to energy loss of partons in the QGP, and well-described by pQCD-based models \cite{Dainese:2004te,Schenke:2009gb,Horowitz:2012cf,Wicks:2005gt}. Recent findings in $p+A$ and $p+p$ collisions, such as strangeness enhancement \cite{ALICE:2013wgn,ALICE:2015mpp}, quarkonium suppression \cite{ALICE:2016sdt}, and collective behavior \cite{CMS:2015yux,ATLAS:2015hzw}, indicate QGP formation. Non-trivial modifications of high-$p_T$ particles have also been observed in small collision systems \cite{ATLAS:2014cpa,PHENIX:2015fgy,ALICE:2017svf}, demanding theoretical explanations.

Applying models that have had successful $A+A$ predictions to $p+A$ and $A+A$ collisions is not straightforward, due to a variety of large system size assumptions within these models. For example, energy loss approaches based on BDMPS-Z \cite{Baier:1996sk} utilize the central limit theorem, which requires a large number of scatterings---a fragile assumption in $A+A$, let alone $p+A$. In this work we consider the removal \cite{Kolbe:2015rvk} of such an assumption---the \emph{large pathlength assumption} $L \gg \mu^{-1}$, where $L$ is the QGP brick length, and $\mu^{-1}$ the Debye screening length---made in the DGLV radiative energy loss model \cite{Gyulassy:2000er,Djordjevic:2003zk}. The correction consists of $\mathcal{O}(e^{- \mu L})$ terms, which were previously assumed to be small, and results in the following novel effects: reduction of energy loss, linear growth with partonic energy, and disproportionate size for incident gluons (cf. usual $C_A / C_F$ color factor scaling). This reduction in energy loss could explain the rapid rise of the charged hadron nuclear modification factor with $p_T$ \cite{Horowitz:2011gd} and the enhancement above unity in $\mathrm{p}+\mathrm{A}$ collisions \cite{Balek:2017man,ALICE:2018lyv}.

In this work we will provide nuclear modification factor $R_{AA}$ and $R_{pA}$ predictions for $A+A$ and $p+A$ collisions, with a focus on the impact of the short pathlength correction to DGLV radiative energy loss on said predictions. We will in particular attempt to, in as reasonable way as possible, make an apples-to-apples comparison of the Wicks-Horowitz-Djordjevic-Gyulassy (WHDG) convolved radiative and collisional energy loss model \cite{Wicks:2005gt}, which has been successful in describing a breadth of leading hadron suppression data, with an energy loss model with the same elastic and geometrical considerations but with a radiative energy loss that includes the short pathlength correction (henceforth the \emph{correction}) as derived in \cite{Kolbe:2015rvk}.

\section{Energy Loss Framework and Results}

The radiative energy loss is calculated according to DGLV \cite{Djordjevic:2003zk} with short path length corrections as derived in \cite{Kolbe:2015rvk}.  The number of radiated gluons $N^g$ differential in the momentum fraction radiated away $x$ is given to first order in opacity $L / \lambda$ by
\begin{align}
  \frac{\mathrm{d} N^g}{\mathrm{d} x}=  \frac{C_R \alpha_s L}{\pi \lambda} \frac{1}{x} \int \frac{\mathrm{d}^2 \mathbf{q}_1}{\pi} \frac{\mu^2}{\left(\mu^2+\mathbf{q}_1^2\right)^2} \int \frac{\mathrm{d}^2 \mathbf{k}}{\pi} \int \mathrm{d} \Delta z \, \rho(\Delta z) \nonumber\\
   \times\left[-\frac{2\left\{1-\cos \left[\left(\omega_1+\tilde{\omega}_m\right) \Delta z\right]\right\}}{\left(\mathbf{k}-\mathbf{q}_1\right)^2+m_g^2+x^2 M^2}\left[\frac{\left(\mathbf{k}-\mathbf{q}_1\right) \cdot \mathbf{k}}{\mathbf{k}^2+m_g^2+x^2 M^2}-\frac{\left(\mathbf{k}-\mathbf{q}_1\right)^2}{\left(\mathbf{k}-\mathbf{q}_1\right)^2+m_g^2+x^2 M^2}\right] \right. 
   \label{eqn:full_dndx}\\
   +\frac{1}{2} e^{-\mu_1 \Delta z}\left(\left(\frac{\mathbf{k}}{\mathbf{k}^2+m_g^2+x^2 M^2}\right)^2\left(1-\frac{2 C_R}{C_A}\right)\left\{1-\cos \left[\left(\omega_0+\tilde{\omega}_m\right) \Delta z\right]\right\}\right. \nonumber
\end{align}
\begin{align}
   \left.\left.+\frac{\mathbf{k} \cdot\left(\mathbf{k}-\mathbf{q}_1\right)}{\left(\mathbf{k}^2+m_g^2+x^2 M^2\right)\left(\left(\mathbf{k}-\mathbf{q}_1\right)^2+m_g^2+x^2 M^2\right)}\left\{\cos \left[\left(\omega_0+\tilde{\omega}_m\right) \Delta z\right]-\cos \left[\left(\omega_0-\omega_1\right) \Delta z\right]\right\}\right)\right],\nonumber
\end{align}
where the first two lines of the above equation are the original DGLV result \cite{Djordjevic:2003zk} and the last two lines are the correction \cite{Kolbe:2015rvk}.  $\omega \equiv x E^+ / 2,~\omega_0 \equiv \mathbf{k}^2 / 2 \omega,~\omega_i \equiv (\mathbf{k} - \mathbf{q}_i)^2 / 2 \omega$, $\mu_i \equiv \sqrt{\mu^2 + \mathbf{q}_i^2}$, and $\tilde{\omega}_m \equiv (m_g^2 + M^2 x^2) / 2 \omega$ following \cite{Djordjevic:2003zk, Kolbe:2015rvk}. Additionally $\rho(\Delta z)$ is the distribution of scattering centers; $\mathbf{q}_i$ is the transverse momentum of the $i^{\mathrm{th}}$ gluon exchanged with the medium; $\mathbf{k}$ is the transverse momentum of the radiated gluon; $\Delta z$ is the distance between production of the hard parton, and scattering; $C_R$ ($C_A$) is the Casimir of the hard parton (adjoint) representation; $\alpha_s$ is the strong coupling, $m_g = \mu / \sqrt{2}$ is the thermal gluon mass; and $M$ is the mass of the hard parton.

Multi-gluon emission is taken into account by assuming gluon emission is independent, leading to a Poisson convolution with \autoref{eqn:full_dndx} as the single-emission kernel, following \cite{Gyulassy:2001nm}. Elastic energy loss is calculated according to \cite{Braaten:1991we} in conjunction with the assumption of applicability of the central limit theorem (large number of elastic scatters), following \cite{Wicks:2005gt, Horowitz:2011gd}. The total probability for energy loss is found as the convolution of the radiative energy loss probability with the elastic energy loss probability, in the same fashion as \cite{Wicks:2005gt}.

We account for path length fluctuations based on the collision geometry by defining an effective path length which depends on the density of the plasma, following WHDG \cite{Wicks:2005gt},
\begin{align}
  L_{\mathrm{eff}}(\mathbf{x}_i, \varphi) &\equiv \frac{1}{\rho_{\text{eff}}} \int_0^{\infty} \mathrm{d} \tau \rho_{\text {part }}(\mathbf{x}_i+\hat{\boldsymbol{\varphi}} \tau)\text{ and } \rho_{\text{eff}} \equiv \left. \int \mathrm{d}^2 x~\rho_{\text{part}}^2(\tau_0, \mathbf{x})  \;\middle/\;\int \mathrm{d}^2 x ~ \rho_{\text{part}}(\tau_0, \mathbf{x}) \right.,
    \label{eqn:effective_length}
\end{align}
where $\rho_{\text{part}}$ is the density of participating nucleons (calculated using profiles from \cite{Schenke:2020mbo}), $\mathbf{x}_i$ is the hard parton production point, and $\boldsymbol{\hat{\varphi}}$ is the direction of propagation. Expansion of the plasma is taken into account via the Bjorken expansion formula $\langle T(\tau) \rangle \approx\left\langle T\left(\tau_0\right)\right\rangle \left[\tau_0 / \tau\right]^{1 / 3} \approx \left \langle T\left( \tau_0 \right) \right\rangle\left[\tau_0 / (L / 2)\right]^{1 / 3}$, where $\langle T(\tau_0) \rangle \propto \rho_{\text{eff}}^{1 / 3}$ and $\mu$ and $\lambda$ are determined dynamically from the temperature (details in \cite{Faraday:2023mmx}). We will compare two distributions of scattering centers $\rho(\Delta z)$: the \emph{exponential distribution} $\rho_{\text{exp.}} (\Delta z) \equiv \frac{2}{L} \exp[ - 2 \Delta z / L]$, and the \emph{truncated step distribution} $\rho_{\text{step}}(\Delta z) \equiv (L-\tau_0)^{-1} \Theta(\Delta z-\tau_0) \Theta(L-\Delta z)$, where $\tau_0 \sim 0.4~\mathrm{fm}$ is the hydrodynamics turn-on time. The exponential distribution is the canonical choice used in WHDG \cite{Wicks:2005gt, Horowitz:2011gd}.
These two choices can be thought of as bounds: the exponential overemphasizes the correction, while the truncated step drastically reduces it by turning off early-time energy loss.

As a first exploration of the effect of the short pathlength correction, we make predictions for both $Pb+Pb$ and $p+Pb$ collisions with $\sqrt{s} = 5.02~\mathrm{TeV}$.

\begin{figure}[!htbp]
  \centering
  \includegraphics[height=5.7cm]{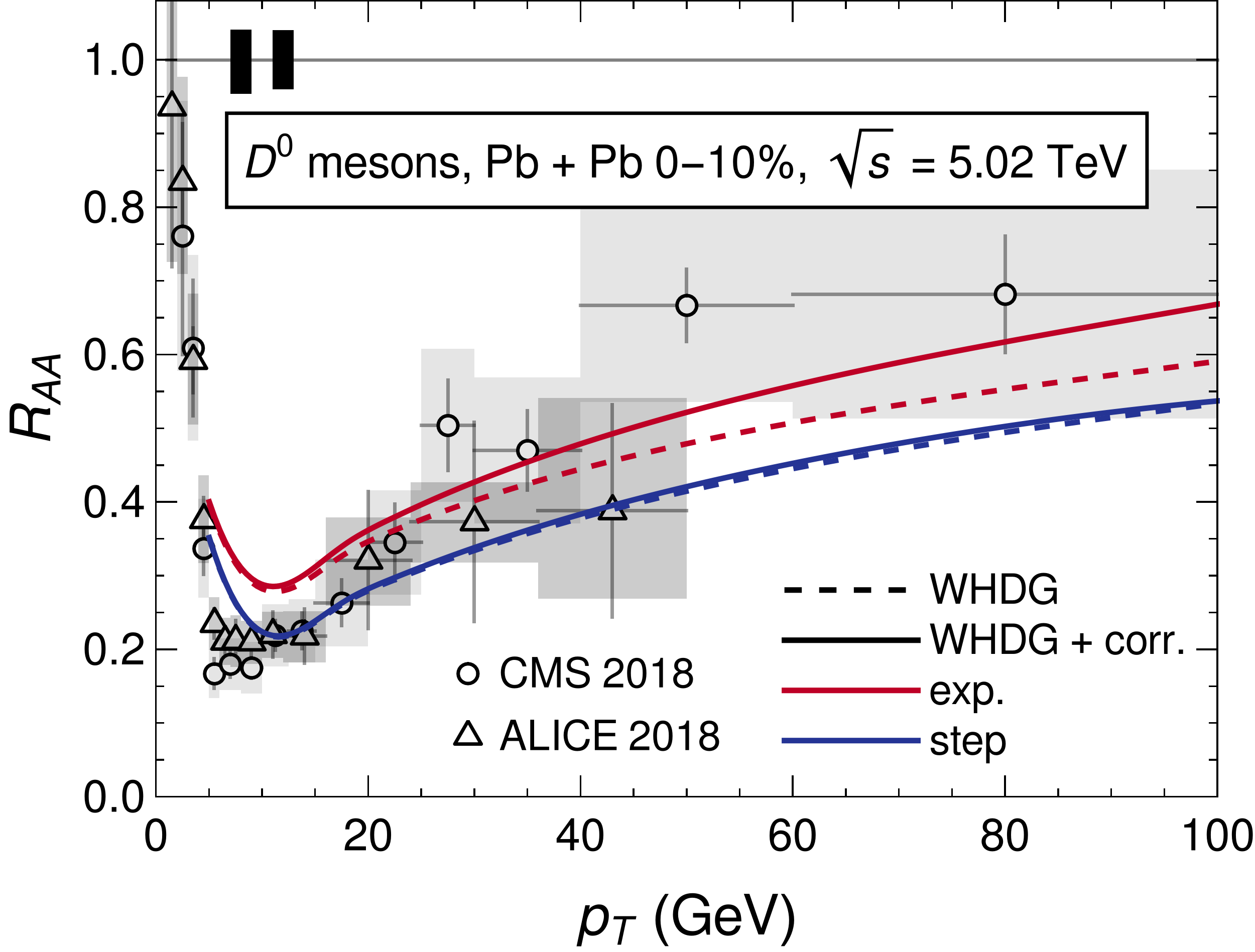}
  \hfill
  \includegraphics[height=5.7cm]{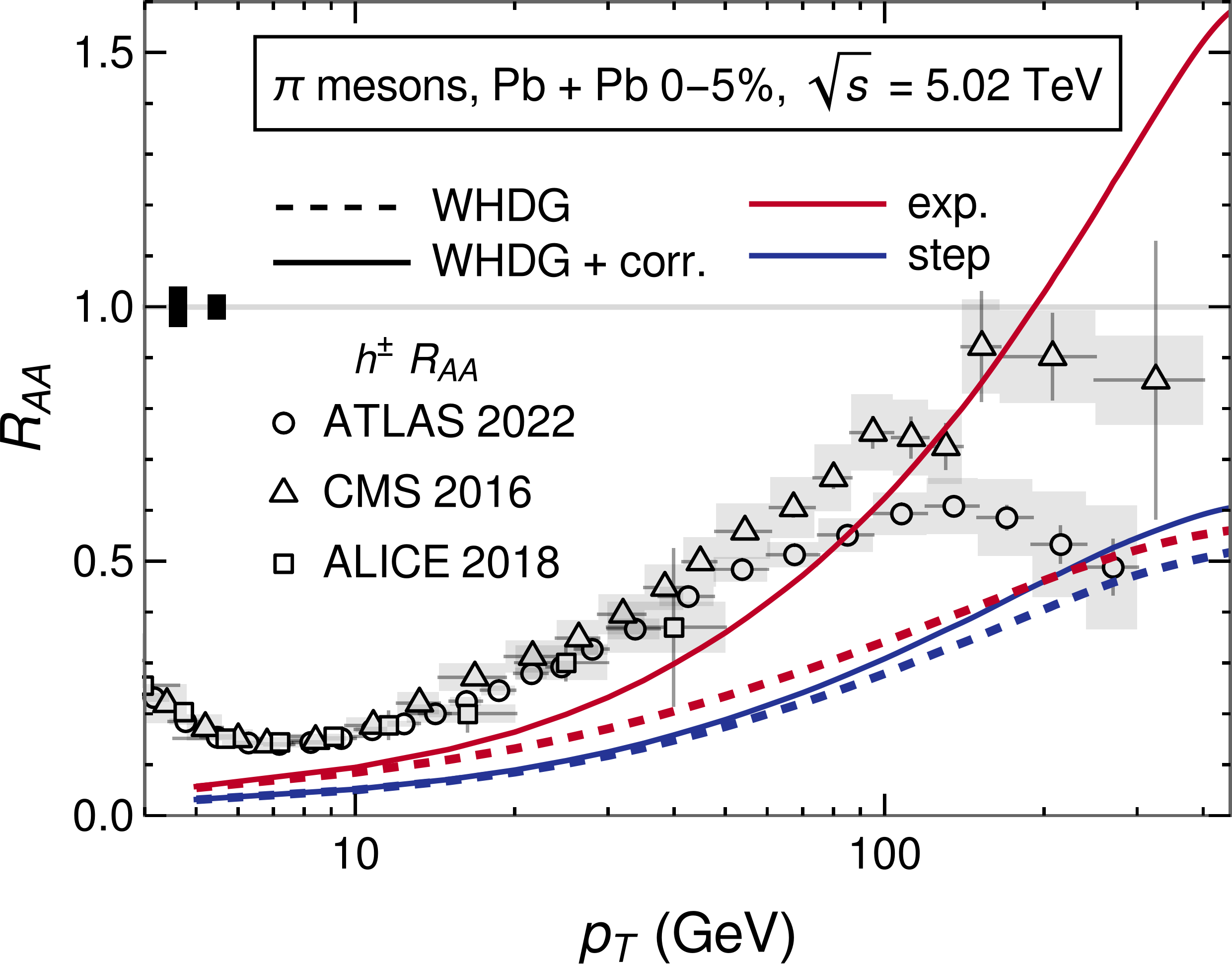}
  \caption{(Left) Plot of the $R_{AA}$ for $D$ mesons as a function of final transverse momentum $p_T$ in $\sqrt{s} = 5.02$ TeV $Pb+Pb$ central 0--10\% (top) and semi-central 30--50\% (bottom) collisions. Data are from ALICE \cite{ALICE:2018lyv} and CMS \cite{CMS:2017uoy}. (Right) Plot of the  $R_{AA}$ for $\pi$ mesons produced in 0--5\% most central $Pb+Pb$ collisions at $\sqrt{s} = 5.02$ TeV, as a function of the final transverse momentum $p_T$. Data from ATLAS \cite{ATLAS:2022kqu}, CMS \cite{CMS:2016xef}, and ALICE \cite{Sekihata:2018lwz}. (Left and Right) Predictions with (solid) and without (dashed) the short path length correction to the radiative energy loss are shown using the exponential (red) and truncated step (blue) distributions for the scattering centers. The experimental global normalization uncertainty on the number of binary collisions is indicated by the solid boxes in the center left of the plot (left to right: CMS, ALICE; CMS, ALICE). In the ATLAS data the normalization uncertainty is included in the systematic uncertainty.}
  \label{fig:RAA}
\end{figure}

\autoref{fig:RAA} shows the predicted $R_{AA}(p_T)$ in central collisions for $D^0$ mesons (left) and for $\pi$ mesons (right), as a function of final transverse momentum $p_T$. For the $D^0$ mesons (left) we find that---for an exponential distribution of scattering centers---the correction is $\lesssim \mathcal{O}(10\%)$, and grows with $p_T$. The exponential vs truncated step distribution is an $\mathcal{O}(10\text{--}20\%)$ effect at high $p_T$, and the truncated step distribution practically removes the correction. We conclude that for heavy flavour mesons in $A+A$ collisions, the correction is negligible.

For the $\pi$ mesons (right), the predicted $R_{AA}(p_T)$ values differ by $\mathcal{O}(10\text{--}20)\%$ between: the exponential distribution without correction, the truncated step distribution with correction, and the truncated step without correction. For the exponential distribution, the correction becomes excessively large, reaching $\mathcal{O}(100\%)$ at $100~\mathrm{GeV}$. For $p_T \lesssim 100~\mathrm{GeV}$ the correction agrees tantalizingly well with data, and possibly could account for the sharp increase in $R^\pi_{AA}(p_T)$ at the LHC; however the asymptotic behaviour of the correction is inconsistent with data. In \cite{Faraday:2023mmx} we perform a thorough analysis of the various assumptions---including collinearity, softness, and large formation time---that are made by both DGLV \cite{Djordjevic:2003zk} and the correction to DGLV \cite{Kolbe:2015rvk}, where we find that the \emph{large formation time} assumption is breaking down at $\mathcal{O}(30\text{--}50)~\mathrm{GeV}$. The breakdown is particularly fast for gluons (which fragment into $\pi$ mesons), indicating that this breakdown could be tied to the excessive correction. This calls for a rederivation of GLV, DGLV, and the correction with the large formation time assumption relaxed.

\begin{figure}[!htbp]
  \centering
  \includegraphics[width=0.48\textwidth]{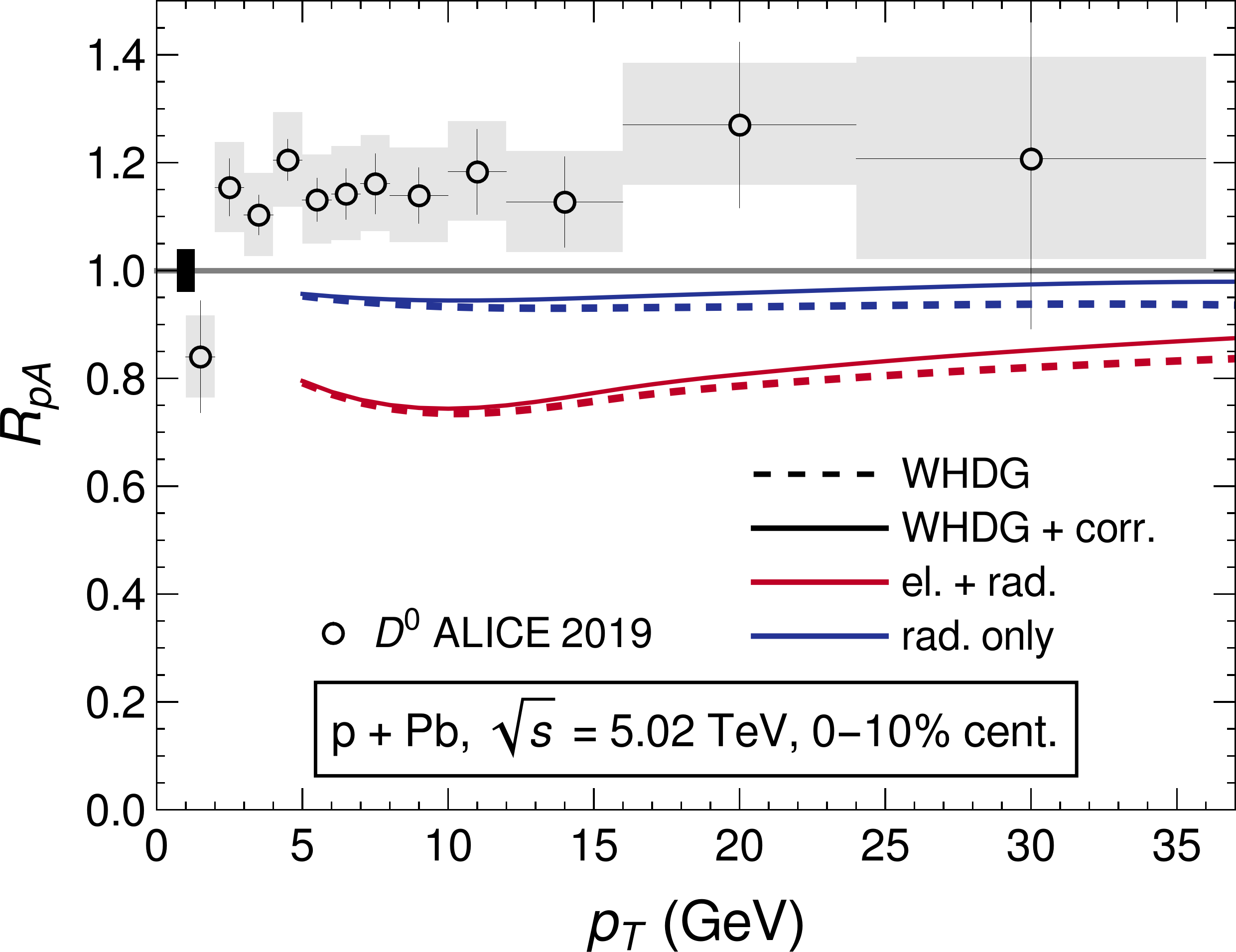}
  \hfill
  \includegraphics[width=0.48\textwidth]{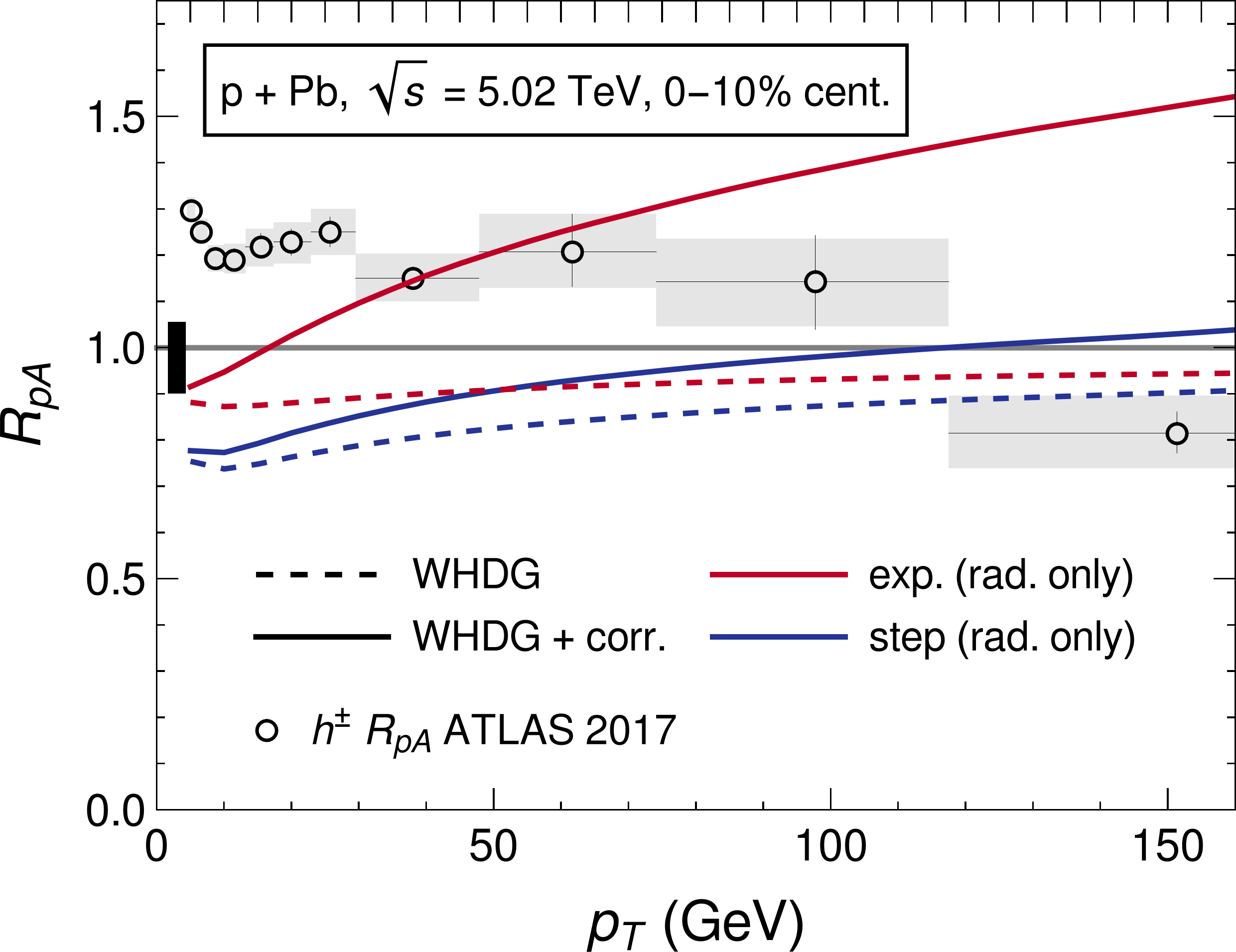}
    \caption{
    (left) The nuclear modification factor $R_{pA}$ for $D^0$ mesons as a function of final transverse momentum $p_T$ in 0--10\% central $p+Pb$ collisions at $\sqrt{s}=5.02$ TeV.  Data are from ALICE \cite{ALICE:2019fhe}. The $R_{pA}$ is calculated both with elastic and radiative energy loss (el. + rad.) and radiative energy loss only (rad. only). (right) The nuclear modification factor $R_{p A}$ for $\pi$ mesons as a function of final transverse momentum $p_T$ in 0--10\% $p+Pb$ collisions at $\sqrt{s}=5.02$ TeV. Only radiative energy loss is included; predictions with (solid) and without (dashed) the correction are shown using the exponential (red) and truncated step (blue) distributions for the scattering centers. Data for charged hadrons are from ATLAS \protect\cite{Balek:2017man}. (left and right) The experimental global normalization uncertainty on the number of binary collisions is indicated by the solid box in the top left corner of the plot.
  }
  \label{fig:RpA}
\end{figure}

\autoref{fig:RpA} shows the nuclear modification factor $R_{pA}$ in central $p+Pb$ collisions for $D^0$ mesons (left) and $\pi$ mesons (right). The left pane of \autoref{fig:RpA} shows corrected (solid) and uncorrected (dashed) $R_{pA}$ for $D^0$ mesons. The $R_{pA}$ was calculated with both convolved elastic and radiative energy loss (red), and radiative energy loss only (blue). Comparing the full elastic and radiative energy loss prediction and the radiative only prediction, it is evident that the elastic energy loss is dominant. This is an artifact of the conventional application of the central limit theorem (CLT) to the number of scatters in the medium. In $\mathrm{p} + \mathrm{A}$ collisions, there are $\sim \!\! 1$ collision, and so CLT breakdown is not surprising. This calls for future short pathlength corrections to elastic energy loss. 

The right pane of \autoref{fig:RpA} shows corrected (solid) and uncorrected (dashed) $R_{pA}$ for $\pi$ mesons with both exponential (red) and truncated step (blue) distributions of scattering centers. The corrected $R_{pA}$ with an exponential distribution shows qualitative agreement with data up to $100~\mathrm{GeV}$, indicating the possibility that measured $R_{pA}>1$ could be---at least in part---due to final state effects. The correction is dramatically reduced when the truncated step distribution is used, indicative of the sensitivity of the correction to $\rho(\Delta z)$.

\section{Conclusions}

We considered the effect of the short pathlength correction to DGLV radiative energy loss, as a component of a sophisticated phenomenological energy loss model based off the WHDG model.  We found that the central limit theorem assumption used in the elastic energy loss results in a dramatic over prediction in suppression in $p+A$ collisions, necessitating short pathlength corrections to elastic energy loss. Furthermore the short pathlength correction to radiative energy loss is excessively large for $R^\pi_{AA}$ and $R^\pi_{pA}$, which may be caused by the breakdown of the large formation time assumption, calling for a short formation time correction. The short pathlength correction is extremely sensitive to the distribution of scattering centers at small distances, requiring more consideration for the most physically reasonable distribution of scattering centers. Our results qualitatively support the possibility of final state effects contributing to the observed $R_{pA}>1$ in experimental data, but quantitative predictions require further theoretical development.

Additional future work in small systems may seek to place energy loss calculations on a more rigorous footing \cite{Clayton:2021uuv} or consider the small system size corrections to thermodynamics \cite{Mogliacci:2018oea}, the equation of state \cite{Horowitz:2021dmr}, and the effective coupling \cite{Horowitz:2022rpp}.

\section*{Acknowledgments}

The authors thank Isobel Kolb\'e for valuable discussions and Chun Shen for supplying hydrodynamic temperature profiles. CF and WAH thank the National Research Foundation and the SA-CERN Collaboration for support.

\bibliography{small_system}

\end{document}